\documentclass[pdflatex,sn-mathphys-num,iicol]{sn-jnl}

\usepackage{graphicx}%
\usepackage{multirow}%
\usepackage{amsmath,amssymb,amsfonts}%
\usepackage{amsthm}%
\usepackage{mathrsfs}%
\usepackage[title]{appendix}%
\usepackage{xcolor}%
\usepackage{textcomp}%
\usepackage{manyfoot}%
\usepackage{booktabs}%
\usepackage{algorithm}%
\usepackage{algorithmicx}%
\usepackage{algpseudocode}%
\usepackage{listings}%

\newcommand{\beq}{\begin{equation}}
\newcommand{\eeq}{\end{equation}}
\newcommand{\bea}{\begin{eqnarray}}
\newcommand{\eea}{\end{eqnarray}}
\newcommand{\nn}{\nonumber}
\newcommand{\eq}{Eq.~}
\newcommand{\eqs}{Eqs.~}
\newcommand{\fig}{Fig.~}

\newcommand{\bx}{{\bf x}}

\newcommand{\bp}{{\bf p}}

\def\lsi{\raise0.3ex\hbox{$<$\kern-0.75em\raise-1.1ex\hbox{$\sim$}}}
\def\gsi{\raise0.3ex\hbox{$>$\kern-0.75em\raise-1.1ex\hbox{$\sim$}}}

\newcommand{\gsim}{\mathop{\gsi}}

\theoremstyle{thmstyleone}%
\theoremstyle{thmstyletwo}%

\theoremstyle{thmstylethree}%

\begin{document}

\title[]{Emergent chiral spin symmetry, non-perturbative dynamics and thermoparticles in hot QCD}


\author[]{\fnm{Owe} \sur{Philipsen}}

\affil[]{\orgdiv{ITP}, \orgname{Goethe-Universit\"at Frankfurt am Main}, \orgaddress{\street{Max-von-Laue-Str. 1},\\ \city{60438 Frankfurt am Main}, \country{Germany}}}


\abstract{Several non-perturbative results for hot QCD are challenging
some aspects of the phase diagram and its associated degrees of freedom which were previously believed to be well understood.
With increasing temperature, the chiral crossover is followed by an intermediate region with an approximate chiral spin symmetry
larger than chiral symmetry, in which pseudo-scalar mesons continue to exist as hadron-like excitations, before at some higher temperature the expected
chiral symmetry is recovered. By testing general formal considerations against lattice data, 
it can be shown that thermally modified versions of stable vacuum particles, so-called thermoparticles, form
the constituents of thermal quantum field theories, with properties quite different from what is expected perturbatively. This ``viewpoint'' aims to raise broader
and, in particular, phenomenological interest in these directions.  }

\keywords{Chiral Symmetry, Confinement, Quark Gluon Plasma, Lattice QCD}



\maketitle

\section{Introduction}\label{sec1}

The QCD phase diagram and the associated degrees of freedom in different regions of temperature and
baryon density are of paramount importance for studies of matter under extreme conditions.
While a severe fermionic sign problem prohibits Monte Carlo simulations of lattice QCD at finite baryon density,
the temperature direction is widely believed to be 
relatively well understood, being fully accessible by lattice simulations, 
functional and, for sufficiently large temperatures, perturbative methods. 
Over the last decades these have produced the prevailing simple picture at zero density, according to which 
QCD undergoes a smooth thermal crossover at $T_\mathrm{ch}\approx 156$ MeV  to a high temperature regime with approximately restored
$SU(2)_L\times SU(2)_R\times U(1)_A$ chiral symmetry 
and dominantly partonic degrees of freedom, i.e., deconfined quarks and gluons. These are expected to interact with still considerable
strength, so as to account for rapid thermalisation and the low viscosity of the resulting near-perfect liquid, which is often called the strongly interacting quark gluon plasma, sQGP. 
\begin{figure*}[t]
\centering
\includegraphics[width=0.32\textwidth]{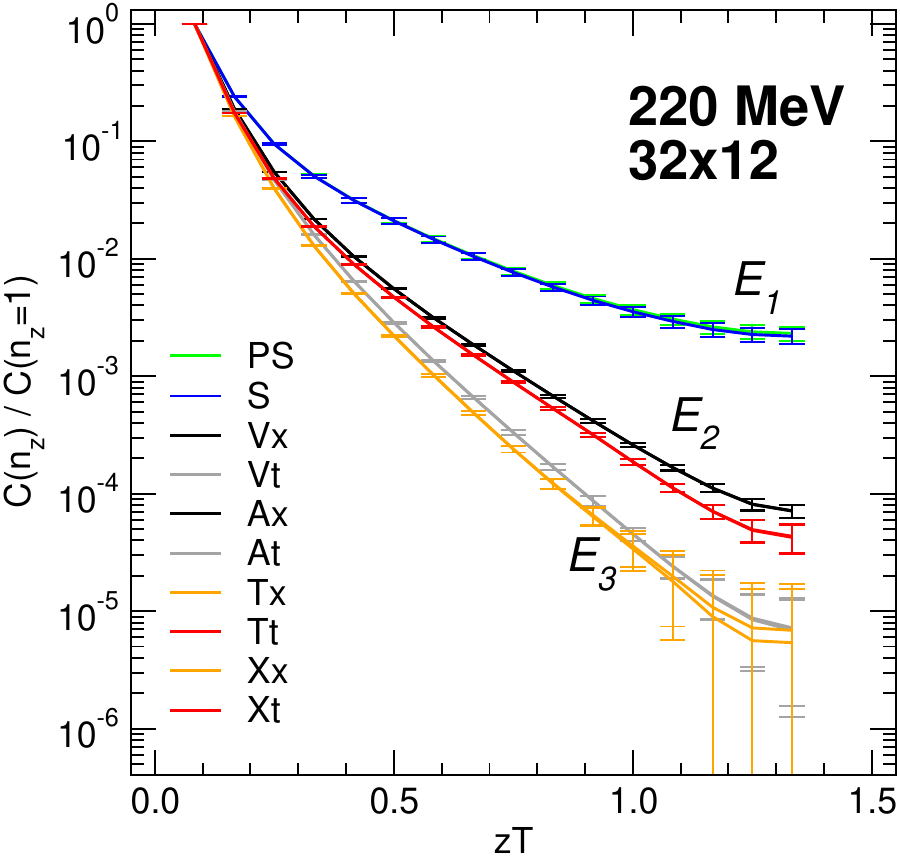}
\includegraphics[width=0.32\textwidth]{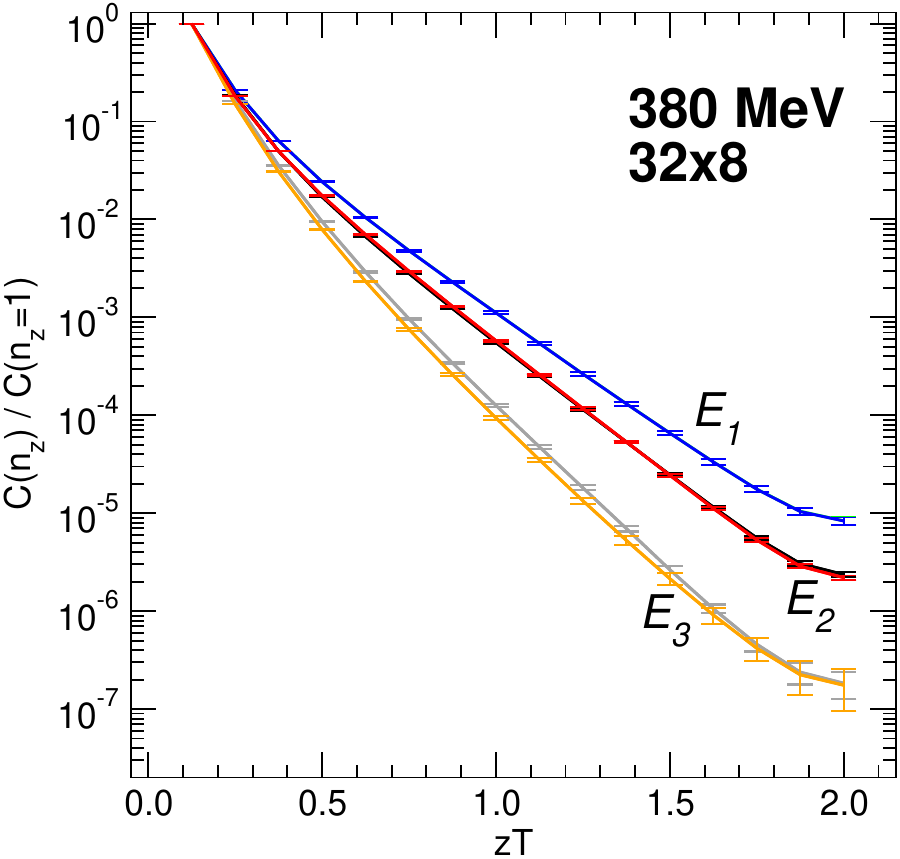}
\includegraphics[width=0.32\textwidth]{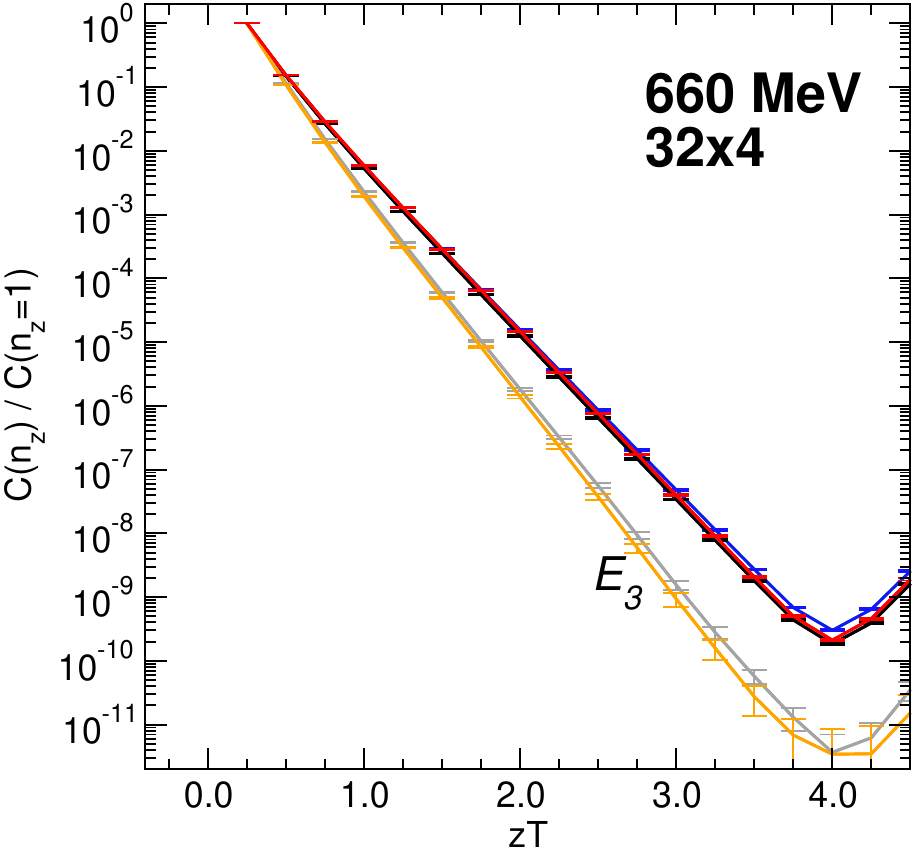}
\caption[]{Spatial correlation functions of QCD with $N_f=2$ quarks with physical light-quark mass. Distinct $E_1,E_2,E_3$ multiplets of the approximate $SU(4)$ chiral spin symmetry, 
 are visible at temperatures above the crossover. 
At large temperatures, these reduce to the multiplets of the ordinary chiral symmetry. From \cite{Rohrhofer:2019qwq}.}
\label{fig:chi_spin1}
\end{figure*}

The purpose of this note is to highlight several theoretical developments, all based on fully non-perturbative results, which are in increasing tension 
with this picture and warrant new phenomenological investigations.
These can be grouped into three categories which complement each other:
\begin{itemize}
\item[\textit{i)}] Investigations of the symmetries realised by QCD show an intermediate regime $T_\mathrm{ch}<T<T_\mathrm{d}$, with an emergent chiral-spin symmetry larger
than the QCD chiral symmetry. Confining  colour-electric flux dominates the dynamics up to some higher temperature $T_\mathrm{d}\sim 2-3 T_\mathrm{ch}$, when the expected
chiral symmetry takes over. 
\item[ \textit{ii)}] Spectral functions of pseudo-scalar mesons show broadened peaks identifiable as thermallly modi\-fied
pion and kaon states in this intermediate regime, suggesting the propagating degrees of freedom to be hadronic.
\item[ \textit{iii)}] General non-perturbative considerations of quantum field theories at finite temperature lead to pronounced differences to vacuum QFT,
which can be verified on the lattice but are not reflected in standard perturbative approaches.
\end{itemize}
The following sections summarise the main results of these developments.

\section{Regions of hot QCD}\label{sec2}

\begin{figure*}[t]
\vspace*{-0.5cm}
\centering
\includegraphics[width=0.4\textwidth]{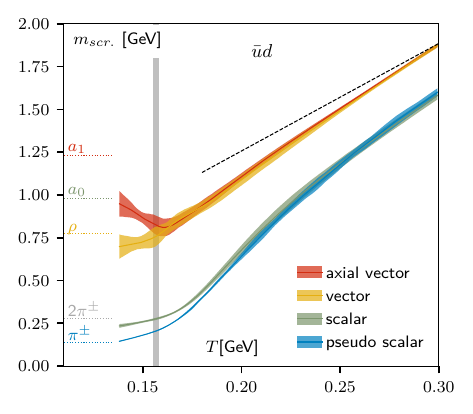}\hspace*{0.5cm}
\includegraphics[width=0.4\textwidth]{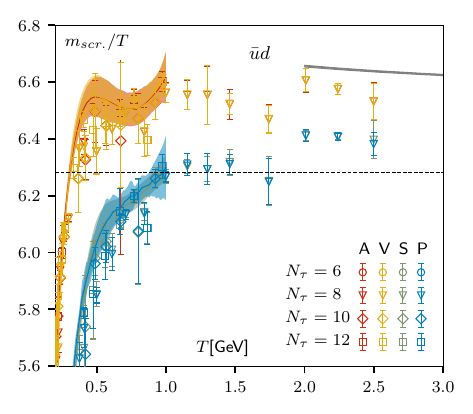}
\caption[]{
Screening masses of the lightest $\bar{u}d$-mesons, evaluated using 
    HISQ fermions. From \cite{Bazavov:2019www}.}
\label{fig:mscr}
\end{figure*}
Phase transitions are usually caused by changes in the symmetries of a system under variation of some external control parameter(s).
In the case of QCD one expects a restoration of the spontaneoulsy broken chiral symmetry with increasing temperature, as well as a change in the confining behaviour, which in
Yang-Mills theory is associated
with a breaking of the global $Z(3)$ center of the $SU(3)$ gauge group. Both symmetries are broken explicitly for finite quark masses, 
so there is no temperature/density regime where
either symmetry is exact. Consequently the would-be order parameters for chiral and center symmetry, the chiral condensate and the Polyakov loop, respectively,
are never zero, but merely change between regions with smaller and larger values. Hence, in physical QCD there is no need for non-analytic phase transitions  between different regions, 
and indeed a thermal analytic crossover is observed in lattice simulations~\cite{Aoki:2006we}.

\subsection{Chiral symmetries}

An $SU(2)_{CS}$ chiral spin transformation of Dirac quark fields is defined by
\bea
\psi(x)&\rightarrow& \exp\Big( i\vec{\Sigma} \cdot\vec{\epsilon}\Big)\psi(x)\quad \mbox{with}\\
\vec{\Sigma}&=&(\gamma_k,-i\gamma_5\gamma_k,\gamma_5)\;,
\quad [\Sigma_i,\Sigma_j]=2i \epsilon_{ijk}\Sigma_k\;. \nn
\eea   
Here $k\in\{0,\ldots 3\}$ labels any of the euclidean gamma matrices. It is apparent that $SU(2)_{CS}\supset U(1)_A$. Furthermore, when combined with 
isospin, $SU(2)_{CS}\times SU(2)_V$ can be embedded into the larger $SU(4)$, 
which contains the usual chiral symmetry of the massless QCD Lagrangian, 
$SU(4)\supset SU(2)_L\times SU(2)_R\times U(1)_A$.

The QCD Lagrangian is not invariant under chiral spin transformations. A thermal medium implies a 
Lorentz frame, and the massless quark action splits into spatial and temporal parts, 
\bea
\bar{\psi}\gamma_\mu D_\mu \psi&=& \bar{\psi}\gamma_0 D_0 \psi + \bar{\psi}\gamma_i D_i\psi\;.
\label{eq:comm}
\eea
The colour-electric part of the quark-gluon interaction is 
CS- and $SU(4)$-invariant, while kinetic terms (and thus the free Dirac action) as well as colour-magnetic interactions are not. 
Chiral spin symmetry therefore is never exact in physical (or even massless) QCD. Nevertheless, its approximate realisation is possible
if the colour-electric quark-gluon interaction dominates the quantum effective action in some dynamical regime. 

There are by now several lattice simulations which demonstrate this to be the case in some temperature window above the chiral crossover.
These work with domain wall fermions with good chiral symmetry, and explicitly identify the appropriate
multiplets pertaining to the enlarged chiral spin symmetry,  both of spatial \cite{Rohrhofer:2019qwq,Chiu:2023hnm,Aoki:2025mue} and 
temporal~\cite{Rohrhofer:2019qal,Chiu:2023hnm} meson correlation functions. An example of spatial correlators is shown in \fig\ref{fig:chi_spin1}.
Of the three multiplets
$E_1$ is due to $U(1)_A$ restoration whereas $E_3$ requires the full chiral symmetry. Both multiplets are expected
by chiral symmetry. What is surprising is the multiplet $E_2$, which does \textit{not} correspond to 
a representation of chiral symmetry, but to one of the larger chiral spin symmetry $SU(4)$. At sufficiently large
temperatures $T>T_d$ the colour-electric flux gets screened and the expected chiral symmetry is recovered. There are then two transitions between three regions,
a low-temperature region with broken chiral symmetry, an intermediate region with an approximate chiral spin symmetry, followed at still
higher temperatures by the expected approximate chiral symmetry. 

For a long time it was argued that the degrees of freedom above $T_\mathrm{ch}$ are partonic, because the Polyakov loop rises through the crossover.
However, the Polyakov loop is not a physical degree of freedom in $N_f=2+1$ QCD, and its expectation value depends on an arbitrary renormalisation procedure.
Moreover, since it represents a static colour source its behaviour is not directly related to the binding of light quarks into hadrons, which has
to be probed by other observables.

Note that the evolution of chiral symmetry continues at still higher temperatures when also the heavy $c,b$-quarks are included. This might be expected on kinematical
grounds, but has not been investigated until very recently.
When quark mass thresholds are sufficiently exceeded, the quark mass splittings become small compared to the leading mesonic
Matsubara modes $\sim 2\pi T$,  resulting
in a hierarchically growing approximate chiral symmetry. This has been demonstrated 
all the way up to  $SU(5)_L \times SU(5)_R\times U(1)_A$  at sufficiently large temperatures~\cite{Chiu:2024jyz,Chiu:2024bqx}.

\subsection{Confinement or Deconfinement?}

Three different regimes for $N_f=2+1$ QCD are also visible \cite{Glozman:2022lda} in mesonic $J=0,1$ screening masses~\cite{Bazavov:2019www},
as shown in \fig\ref{fig:mscr}.
On the left
one observes chiral symmetry restoration, with screening masses becoming degenerate at $T_\mathrm{ch}$, as expected (the degeneracy of
the pion with the two-pion state instead of the $a_0$ is a known artefact of the staggered taste 
splitting).
Another marked change of dynamics is visible in the right plot, where $m_{scr}/T$ changes from nearly vertical to
almost horizontal behaviour within a narrow temperature range. At its high temperature end, all screening masses overshoot the leading perturbative  $\sim 2\pi T$ 
and display a sizeable spin effect.
But the horizontal part of the plot is qualitatively compatible with the perturbative form
\begin{eqnarray}
\frac{m_{scr}}{2\pi T}=1+p_2 g^2(T) + p_3 \, g^3(T)+p_4 g^4(T)\;,
\label{eq:psv}
\end{eqnarray}
with a logarithmic $T$-dependence through the running coupling only and constants $p_i$. This changes abruptly between $T_d\sim 500$ MeV for
all four quantum number channels shown in \fig\ref{fig:mscr}. The sudden change in $T$-dependence is not 
compatible with \eq(\ref{eq:psv}), nor can it be accomodated by higher order corrections. 
The same abrupt bending is observed across 12 different quantum number channels~\cite{Bazavov:2019www}, and suggests a change to non-perturbative dynamics as $T<T_d$. 

In the intermediate temperature range, the approximate realisation of chiral spin symmetry 
implies dominance of the colour-electric quark gluon interaction, as discussed above. Since this interaction 
produces the confining flux between quarks and anti-quarks, one would expect confinement to persist in this regime. This is borne out 
by an increasing number of non-perturbative studies using different observables. In particular, colour-electric flux tubes between static 
quark anti-quark pairs inserted in a dynamical QCD medium with $N_f=2+1$ quarks with physical masses are found to be thermally modified, but do persist 
above the chiral crossover~\cite{Baker:2025bja}. A deconfinement transition at some $T_\mathrm{d}>T_\mathrm{ch}$ is also seen in full QCD by the behaviour of 
gauge field structures like magnetic monopoles~\cite{Cardinali:2021mfh} and center vortices~\cite{Mickley:2024vkm}. Since no exact symmetries are involved, this
second transition is presumably also a crossover. The value of $T_d$ depends strongly on the number of quark flavours, their masses and the observable used to define it.
It would be most valuable to work out any connections between these different observables and investigations.

\begin{figure}[t]
\vspace*{-0.5cm}
\centering
\includegraphics[width=0.4\textwidth]{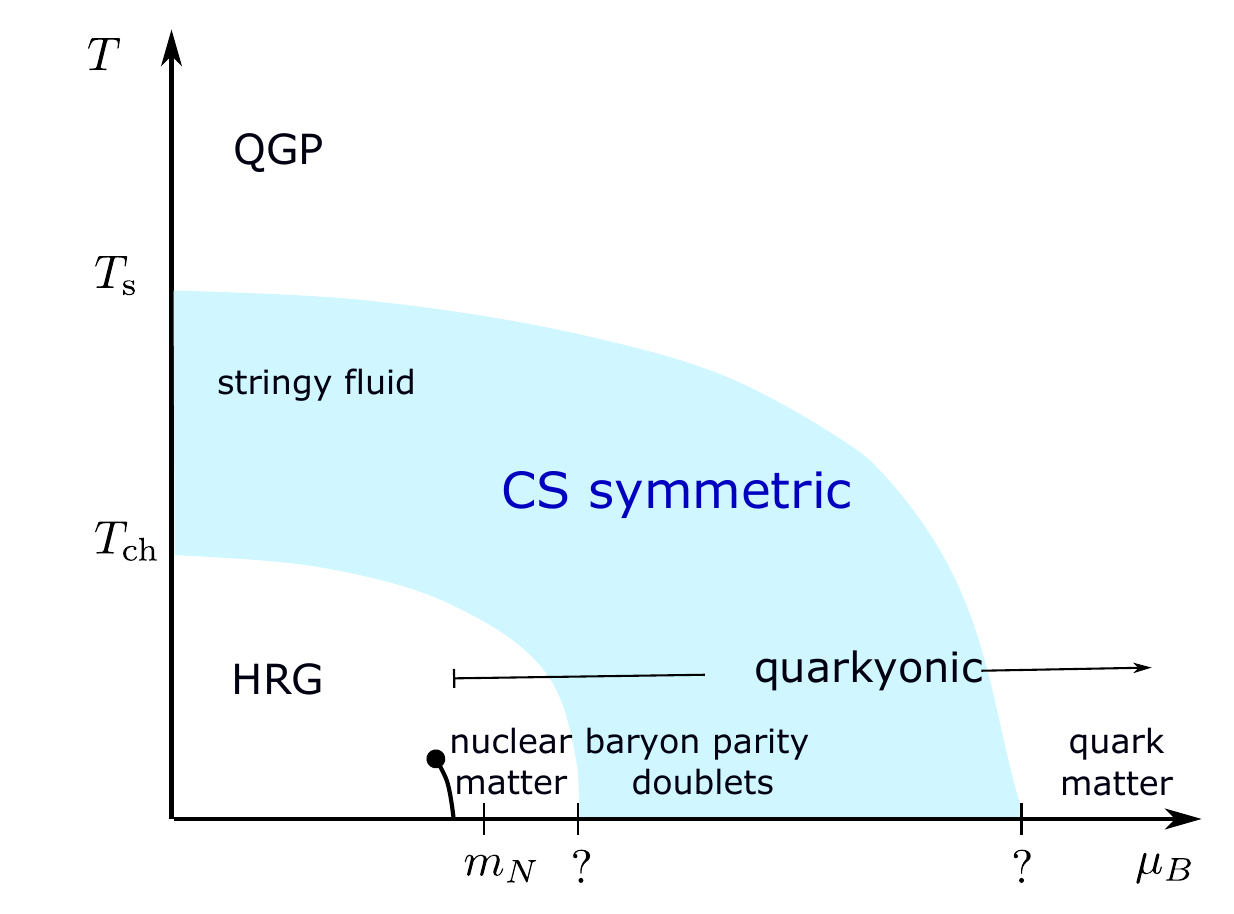}
\caption[]{
Possible QCD phase diagram with a band of approximate chiral spin symmetry (its extension to $\mu_B/T\gsim 1$ is not known).
A chiral phase transiton line with endpoint may be near its lower boundary.
From \cite{Glozman:2022lda}.}
\label{fig:pd}
\end{figure}

Note that a baryo-chemical potential term in the action, $\mu_B/3\;\bar{\psi}\gamma_0\psi$,  
is invariant under chiral spin symmetry. The chiral-spin symmetric window at zero density must therefore continue as a band to $\mu_B\neq 0$. 
At least for $\mu_B/T\lsi 3$ one can infer from chiral symmetry and the behaviour of screening masses
at imaginary chemical potential that the boundaries of the band bend downwards~\cite{Glozman:2022lda}. It is then plausible for it to connect
to a cold and dense quarkyonic regime, which is also expected to be chirally symmetric and confining \cite{McLerran:2007qj}, such as sketched in \fig\ref{fig:pd}. 
However, with 
no fully controlled finite density calculations available, this remains an interesting speculation so far. 

\section{Spectral functions and thermoparticles}

\begin{figure*}[t]
\vspace*{-0.5cm}
\centering
\includegraphics[width=0.45\textwidth]{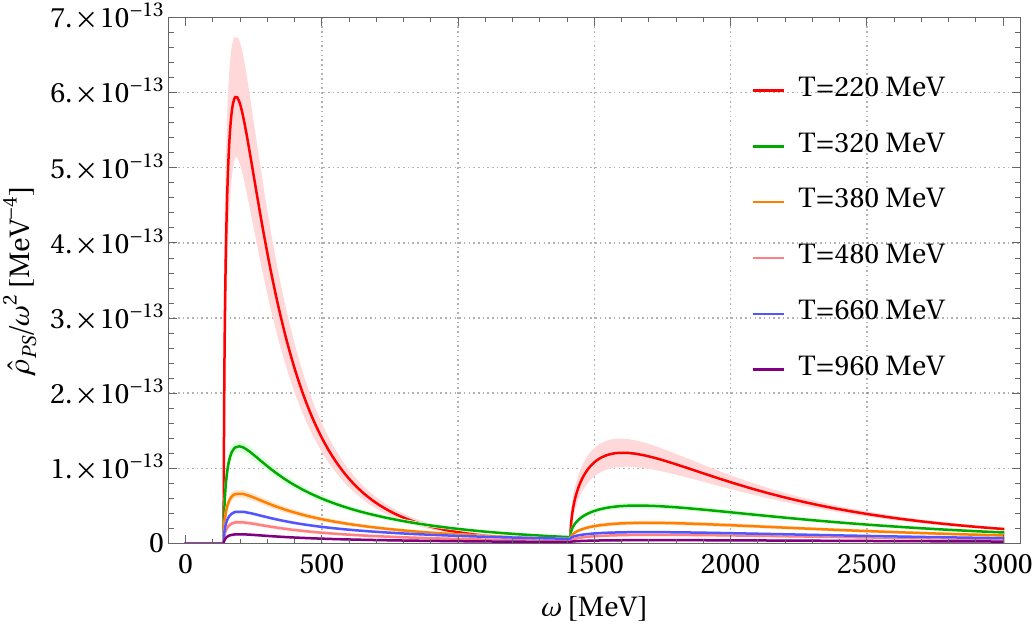}\hspace*{0.5cm}
\includegraphics[width=0.45\textwidth]{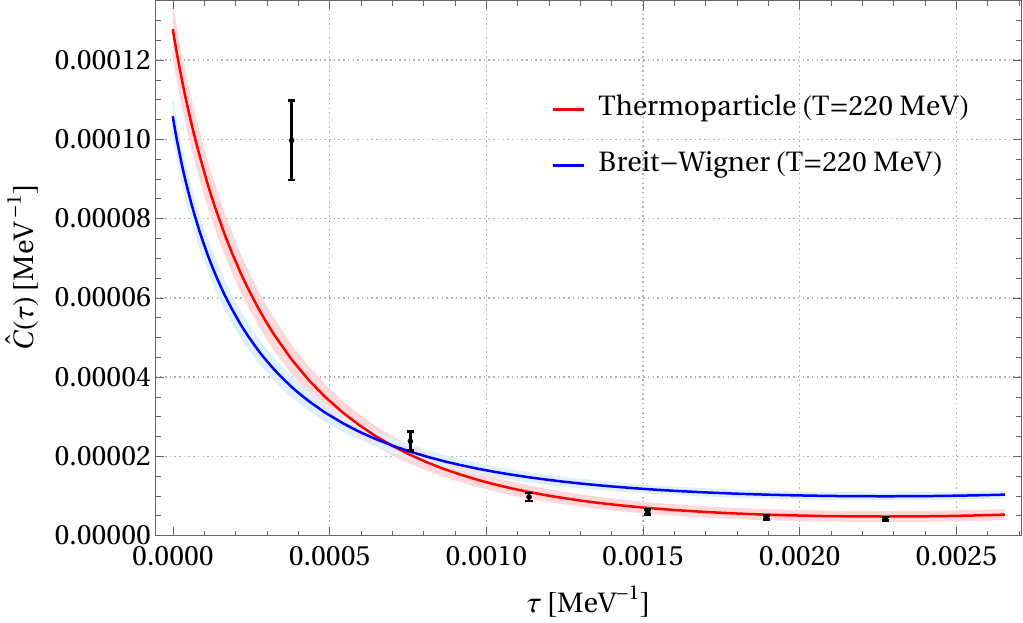}
\caption[]{Left: Pion spectral function reconstructed from the correlators in  \cite{Rohrhofer:2019qwq}. 
Right: Temporal correlator predicted by a thermoparticle (red) or Breit-Wigner (blue) spectral function, 
compared to lattice data  from \cite{Rohrhofer:2019qal}. Figures modified from \cite{Lowdon:2022xcl}.
\label{fig:spec} }
\end{figure*}
The most direct access to the degrees of freedom which dominate correlators of gauge-invariant operators is offered by spectral functions.
These obey relativistic microcausality, as in vacuum. But Lorentz symmetry is broken by the temperature, constraining meson correlators
to be periodic in  euclidean time (the KMS condition).
These properties imply a rigorous representation of the spectral function~\cite{Bros:1998ua,Bros:1996mw,Nair:2025jgl} 
\bea
\rho_{\text{PS}}(\omega,\bp) = \int_{0}^{\infty} \! ds \int \! \frac{d^{3}u}{(2\pi)^{2}}&& \nn \\
& & \hspace*{-4cm}\times \epsilon(\omega) \, \delta\!\left(\omega^{2} - (\bp-\mathbf{u})^{2} - s \right)\widetilde{D}_{\beta}(\mathbf{u},s)\;,
\label{eq:commutator_rep}
\eea
with $\beta=1/T$ and the thermal spectral density $\widetilde{D}_{\beta}(\mathbf{u},s)$. Note that the correct 
K\"allen-Lehmann vacuum representation is attained smoothly as $T\rightarrow 0$.  
In an isotropic medium spatial correlators and thermal spectral density  
are related by~\cite{Lowdon:2022xcl}
\begin{align}
C(z) = \frac{1}{2}\int_{0}^{\infty} \! ds \int^{\infty}_{|z|} \! dR \ e^{-R\sqrt{s}} D_{\beta}(R,s).
\label{C_int}
\end{align}
For stable massive particles like
QCD pions, $\widetilde{D}_{\beta}(\mathbf{u},s)$ is reported to split into discrete particle and continuous scattering contributions~\cite{Bros:2001zs},  
\begin{align}
\widetilde{D}_{\beta}(\mathbf{u},s)= \widetilde{D}_{m,\beta}(\mathbf{u})\, \delta(s-m^{2}) + \widetilde{D}_{c, \beta}(\mathbf{u},s)\;.
\label{eq:decomp}
\end{align} 
The damping factor $\widetilde{D}_{m,\beta}(\mathbf{u})\rightarrow (2\pi)^3\delta^3(\mathbf{u})$ for $T\rightarrow 0$, and the first term reduces to the known on-shell particle 
contribution in vacuum. The damping factor contains all temperature effects a particle experiences in medium via multiple scattering, smoothing the delta function
to finite height and width. This term describes a so-called thermoparticle, representing a constituent of the medium~\cite{Buchholz:1993kp}.
The second term contains all contributions from multi-particle scattering states, including Landau damping.  

At large distances $z$ and not too high temperatures, the spatial correlator in \eq(\ref{C_int}) is dominated by the first term 
in \eq(\ref{eq:decomp})~\cite{Bros:2001zs}. 
Neglecting the continuum part entirely in a first approximation, the extraction of the spectral function becomes straightforward without any ill-posed inversion problem.
In \cite{Lowdon:2022xcl}, this was applied to the pion correlators from \cite{Rohrhofer:2019qwq} shown in \fig\ref{fig:chi_spin1}. 
First, the spatial pion correlators are fitted by the sum of two exponentials representing the $\pi$ and its first excitation, $\pi^*$.
This provides 
the damping factors 
\beq
D^{\pi,\pi^*}_{m,\beta}(|\bx|)=\alpha_{\pi,\pi^*}\exp(-\gamma_{\pi,\pi^*} |\bx|),
\eeq 
from which the spectral function can be reconstructed 
using~\eqs(\ref{eq:commutator_rep}), the first term of (\ref{eq:decomp}) and the corresponding vacuum masses $m_\pi,m_{\pi^*}$.
The result is shown in \fig\ref{fig:spec} (left) and displays the vacuum particle thresholds followed by a pronounced
peak structure for both the pion and its first excitation. As the temperature increases, the peaks widen 
and gradually disappear into a continuum consistent with a multi-quark scattering state. However, this happens at temperatures
significantly above $T_\mathrm{ch}$, while in the chiral-spin symmetric window the pion remains a non-perturbative, hadron-like
excitation. A similar observation is made for all other flavour combinations of light pseudo-scalar mesons in $N_f=2+1$ QCD~\cite{Bala:2023iqu}.
There is an important and highly constraining quality check one can perform for \textit{any} reconstructed spectral function: in order to be correct
it must describe both spatial \textit{and} temporal correlators. The hadron-like pseudo-scalar spectral functions given in \cite{Lowdon:2022xcl,Bala:2023iqu} 
pass this test, as shown for one example in \fig\ref{fig:spec} (right). From this we can conclude that, at least for not too high temperatures, the correlation functions
are do\-minated by the thermoparticle contributions, while multi-meson scattering states and Landau damping are not yet quantitatively important.
This is of course expected to change as the temperature is further increased.   

The finding that the pion survives the chiral crossover is further supported by considering the chiral limit, where the pion is a true Goldstone boson.
Unfortunately, massless QCD cannot be simulated directly. But in view of the generality of the Goldstone theorem, the finite temperature features of Goldstone bosons 
should be qualitatively the same in all theories to which the theorem applies.   
One can then study  a simple $U(1)$ model of
complex scalar fields. In its broken phase this model features a truly massless Goldstone boson, and the symmetry gets thermally restored through a 
second order phase transition rather than a crossover. A detailed investigation of high precision lattice correlators shows that, also in this case, the 
Goldstone mode behaves as a (massless) thermoparticle, which can be clearly identified on \textit{both} sides of the thermal phase transition, albeit
with modifed properties due to the thermal damping factor~\cite{Lowdon:2025fyb,Lowdon:2025ait}.        

\section{Inconsistencies of thermal perturbation theory}

\begin{figure*}[t]
\vspace*{-0.5cm}
\centering
\includegraphics[width=0.5\textwidth]{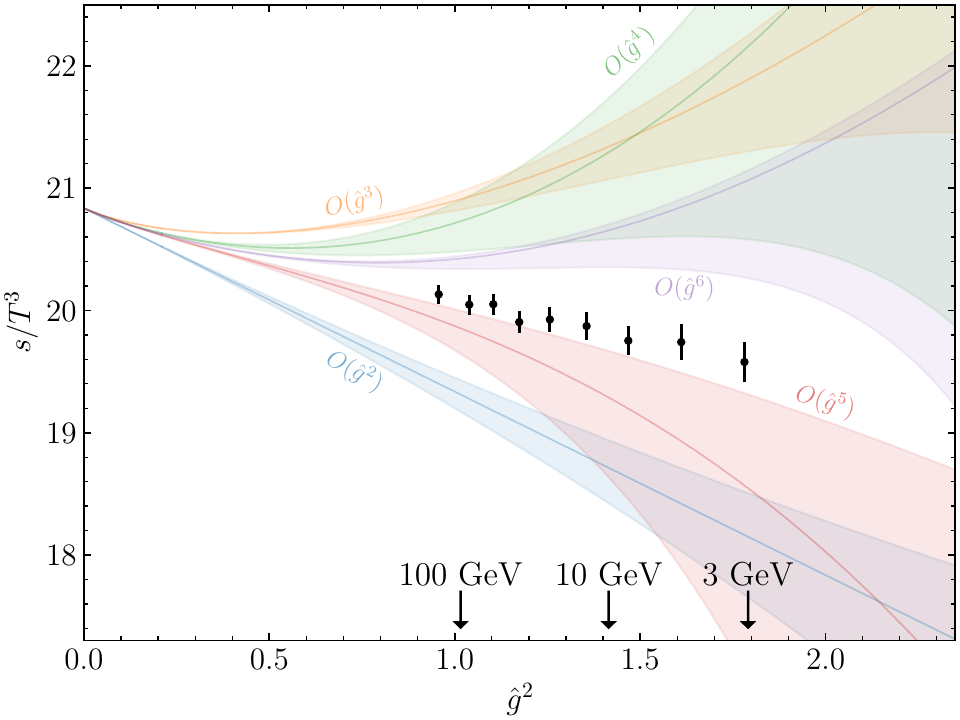}\hspace*{0.2cm}
\includegraphics[width=0.5\textwidth]{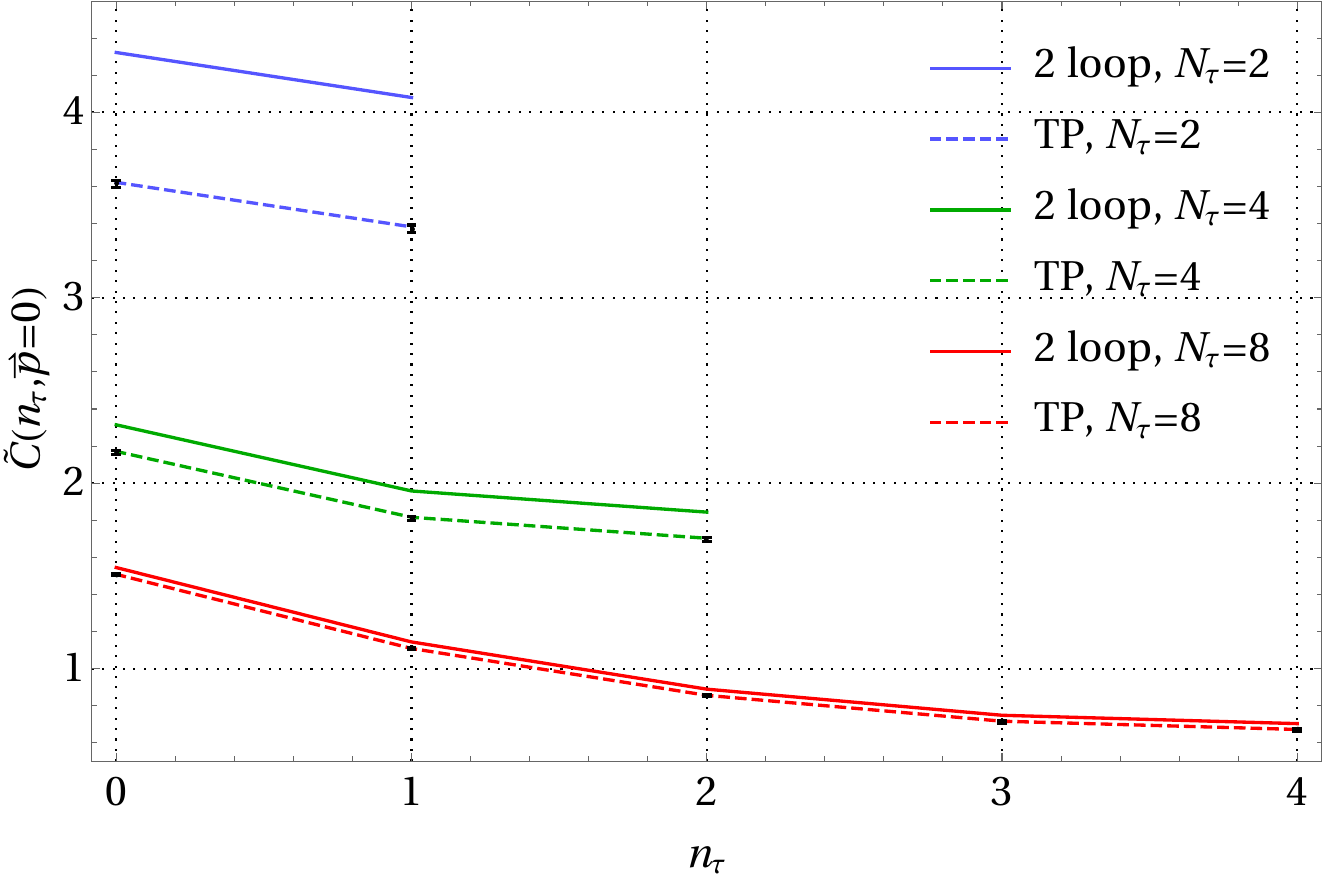}
\caption[]{Left: Entropy density evaluated on the lattice~\cite{Bresciani:2025mcu} compared to perturbation theory \cite{Kajantie:2002wa}, from~\cite{Bresciani:2025mcu}. 
Right: Temporal correlator in $\phi^4$-theory of a massive real scalar field, 
compared to perturbation theory and the thermoparticle contribution extracted from the corresponding spatial correlators. From \cite{Lowdon:2022xcl}.
\label{fig:pert} }
\end{figure*}
Existing expertise about finite $T$ quantum field theories has largely been shaped by perturbation theory. The general belief is that this is guaranteed to work 
for sufficiently high temperatures, with hopefully only quantitative modifications at lower temperatures, in line with the picture of the sQGP. 
However, increasingly accurate lattice data over increasing parameter ranges challenge this expectation.

While it is well known that finite temperature perturbation theory converges more slowly than its zero temperature counter part, the full severity of this 
problem is illustrated by a recent comparison with high temperature lattice results, \fig\ref{fig:pert} (left). Even at $T\sim 100$ Gev there is no sign of rapid convergence of 
the series. 
The standard explanation for this misery are the infrared singularities encountered by perturbation theory with massless particles like gluons. 
Attempts to resolve these problems include the use of effective field theory~\cite{Braaten:1994na}, infinite resummations~\cite{Karsch:1997gj,Andersen:2000yj,Andersen:2008bz,Chiku:1998kd,Braaten:1989mz} and variational techniques like the two-particle irreducible (2PI) formalism~\cite{Blaizot:2000fc,vanHees:2001ik,Blaschke:2013zaa}. 

However, \fig\ref{fig:pert} (right) shows a similar problem for the two-point correlation function in $\phi^4$-theory of a real, massive scalar  field. There is no 
confinement in this theory and no infrared divergence. While perturbation theory is perfectly accurate
on the coldest $N_\tau=(aT)^{-1}=8$ lattice, it increasingly deviates from the full temporal correlator as the temperature rises to $N_\tau=2$. This suggests
a more fundamental problem with the perturbative approach at finite temperature. 
Indeed, it has been pointed out a long time ago that perturbation theory about \textit{free particles} cannot be applied to the finite temperature situation, 
see e.g.~\cite{Narnhofer:1983hp,Landsman:1988ta,Weldon:2001vt}.    
The physical reason is that, in the thermodynamic limit and equilibrium, the medium is everywhere at all times, ruling out the existence of free
on-shell particle states even at asymptotically large times or distances. Formally, this strictly precludes states with real dispersion relations from 
interacting theories~\cite{Narnhofer:1983hp}. Using free states as starting point for the perturbative series leads to branch points overlapping with 
the vacuum pole structure, and hence inconsistencies in the renormalisation~\cite{Weldon:2001vt}. 

By contrast, the dominant states at asymptotically large times in scalar field theories can be shown to be the thermoparticle states \cite{Bros:2001zs},
representing the constituent particles modified by thermal damping. While no analytic solution exists for the asymptotic fields of real scalar $\phi^4$-theory with positive coupling, 
the dotted lines
going through the data points in \fig\ref{fig:pert} (right) are the prediction of the thermoparticle contribution extracted from the spatial lattice correlator, which perfectly reproduces
both correlators at all temperatures studied.

Besides accuracy issues, there are important qualitative aspects concerning the interpretation of the physical degrees of freedom. 
Textbook resummed perturbation theory predicts for the real scalar field a plasmon excitation with a finite width of Breit-Wigner shape~\cite{Wang:1995qf}.
By contrast, thermoparticles are stable constituents, i.e.~thermally modified particles with a vacuum limit. These appear to be dominant at least at moderate temperatures. 
For the pion, its thermoparticle nature would imply a (retarded/advanced) propagator of the form
($k_0=p_0\pm i\epsilon$)~\cite{Lowdon:2022ird}  
\begin{align}
\widetilde{G}_{\beta}^{\pi}(k_{0},\vec{p}) =\frac{\alpha}{|\vec{p}|^{2}-k_{0}^{2}+m_{\pi}^{2}+\gamma_\pi^{2}+2\gamma\sqrt{m_{\pi}^{2}-k_{0}^{2}}}. \nn
\label{propagator_pion}
\end{align}
Because of the square root in the denominator, there are branch points and the corresponding correlator is \textit{not} dominated by simple poles, which is the usual
assumption required for effective theory descriptions of soft pions\footnote{I thank  P.~Lowdon for pointing this out to me.}, see e.g.~\cite{Son:2001ff}.

\section{Conclusions}

Lattice simulations of quantum field theories are now producing quantitatively accurate results for
QCD at physical quark masses and finite temperature, as well as for different model systems of interest, 
with some unexpected features. In particular, 
there is an intermediate temperature
region with a dynamically generated, approximate chiral spin symmetry, which is larger than the standard chiral symmetry and appears to exhibit confining dynamics. In
mesonic spectral functions, pions and kaons are clearly recogniseable as  thermoparticles with medium-modified properties (thermal broadening) at temperatures 
above the chiral crossover. The thermoparticle concept also gives a successful description of massive real scalar particles in $\phi^4$-theory and massless Goldstone modes for a scalar $U(1)$ theory. It thus appears to be a universal feature of quantum field theories at finite temperature. 
Thermoparticle propagators have an intricate singularity structure and are not dominated by simple poles, with 
repercussions for effective theory descriptions.  It would be most interesting to work out experimental consequences of these theoretical developments.\\

\noindent
{\bf Acknowledgements}: I thank Peter Lowdon for continued collaboration, numerous discussions and comments on the manuscript.
This work was partially supported by the Deutsche Forschungsgemeinschaft (DFG, German Research Foundation) through the CRC-TR 211 'Strong-interaction matter under extreme conditions'- project number 315477589 - TRR 211. 


\bibliography{references}

\end{document}